\documentstyle[aasms4]{article}

\lefthead{}
\righthead{}

\def\cm3{{\rm ~cm}^{-3}}
\def\ltsima{$\; \buildrel < \over \sim \;$}
\def\ltsim{\lower.5ex\hbox{\ltsima}}

\begin{document}
\title{Meson Synchrotron Emission from Central Engines of \\
Gamma-Ray Bursts with Strong Magnetic Fields}

\author{Akira Tokuhisa$^{1,2,4}$, Toshitaka Kajino$^{1-4}$}

\altaffiltext{1}{Theoretical Astrophysics Division, National
Astronomical Observatory, Mitaka, Tokyo, 181-8588 Japan;
tokuhisa@th.nao.ac.jp, kajino@th.nao.ac.jp}
\altaffiltext{2}{Department of Astronomy, School of Science,
University of Tokyo, Bunkyo-ku, Tokyo, 113-0033 Japan}
\altaffiltext{3}{Department of Astronomical Science, Graduate 
University for Advanced Studies, Mitaka, Tokyo, 181-8588 Japan}
\altaffiltext{4}{Institute for Nuclear Theory, 
University of Washington, Box 351550, Seattle, WA 98195, U.S.A.}

\begin{abstract}
Gamma-ray bursts (GRBs) are presumed to be powered by still unknown 
central engines for the timescales in the range $1ms \sim$ a few $ s$.
We propose that the GRB central engines would be a viable site for
strong meson synchrotron emission if they were the compact astrophysical
objects such as neutron stars or rotating black holes with extremely strong 
magnetic fields $H \sim10^{12} - 10^{17}G$ and if protons or heavy nuclei were
accelerated to ultra-relativistic energies of order $\sim 10^{12}-10^{22}eV$.  
We show that the charged scalar mesons like $\pi^{\pm}$ and heavy vector mesons
like $\rho$, which have several decay modes onto $\pi^{\pm}$, could be emitted
with high intensity a thousand times larger than photons through strong 
couplings to ultra-relativistic nucleons. These meson synchrotron emission 
processes eventually produce a burst of very high-energy cosmic neutrinos
with $10^{12} eV \leq E_{\nu}$. These neutrinos are to be detected during the
early time duration of short GRBs.  

\end{abstract}


\section{Introduction}
Accumulating observational evidence shows the exsistence of astrophysical
objects with extremely strong magnetic fields $\sim10^{15}G$.
Kouveliotou et al. (1998) have revealed that a galactic X-ray pulsar 
with an estimated magnetic field of $\sim8\times10^{14}G$ causes
recurrent bursts of soft $\gamma$-rays, which are called 
soft gamma repeaters.
Gamma-ray bursts (GRBs) with short and more intense bursts of
$100keV\sim1MeV$ photons still remain puzzling
although it has been progressively clearer that they are
likely to have a cosmological origin.
Some of the theoretical models of GRBs
(Usov 1992, Klu\'{z}niak \& Ruderman 1998, and Pacsy\'{n}ski 1998)
invoke compact objects for their central engines 
such as neutron stars or rotating black holes
with extremely strong magnetic fields $H\sim10^{16\sim17}G$ in 
order to produce ultra-relativistic energy flow of a huge Lorentz factor 
$\Gamma\sim10^{3}$.
Furthermore, neutron stars or black holes with strong magnetic fields
have been proposed by many authors (Hillas 1984, and references therein) 
as an acceleration site of ultra high-enrgy cosmic rays (UHECR).
Milgrom \& Usov (1995) suggested possible association of the two 
highest energy UHECRs with strong GRBs in their error boxes.

An ultra-relativistic nucleus gives rise to efficient meson emission,
in analogy with the canonical photon synchrotron radiation,
in such a strong magnetic field because it couples strongly 
to meson fields as well as to an electromagnetic field. 
Due to its large coupling constant the meson synchrotron emission is 
$\sim10^{3}$ times stronger than the usual photon synchrotron radiation.
Ginzburg \& Syrovatskii (1965a, b) calculated the intensity of 
$\pi^{0}$ emission by a proton in a given magnetic field. 
At that time, however, they could hardly find the astrophysical sites 
to which their formulae could be applied.

In the present paper we propose that GRB central engines could be 
a viable site for strong meson emission.
Waxman \& Bahcall (1997) proposed that high energy neutrinos with 
$\sim10^{14}eV$ are produced by photomeson interactions on 
shock-accelerated protons in the relativistic fireball
(Rees \& M\'{e}sz\'{a}ros 1992). 
Pacsy\'{n}ski \& Xu (1994) suggested that charged pions, 
which are produced in pp collisions when the kinetic energy
of the fireball is dissipated through internal collisions within
the ejecta, produce a burst of $\sim10^{10}eV$ neutrinos.
Our proposed process of meson production is different from these two
which can operate without magnetic fields.
The very rapid variability time scale, $\sim0.1ms$, of many GRBs implies
that each sub-burst reflects the intrinsic primary energy release 
from the central engine (Sari \& Piran 1997).  Thus, the length 
scale of the central engine is estimated to be $\sim10km$.  
The BATSE detector on board of Compton Gamma-Ray
Observatory found the shortest burst of a duration of 5ms 
with substructure on scale of 0.2ms (Bhat et al. 1992). 
Accumulated BATSE data (Fishman \& Meegan 1995) confirmed that 
more than 25\% of total events are short ($\le 2s$) bursts.
Therefore, if the central engines of GRBs were the compact stellar
objects like neutron stars or rotating black holes associated
with strong magnetic fields, relativistic protons or
heavy nuclei would trigger meson synchrotron emission
whose decay products could provide several observational signals
even before the hidden explosion energy is transported to 
the radiation in the later stage of a relativistic fireball.

We extend the previous treatment of $\pi^{0}$ emission of
Ginzburg \& Syrovatskii (1965a, b) to several kinds of neutral 
and charged mesons, which couple to a nucleon 
through scalar or vector type interaction,
in a manner somewhat different from theirs. 
If the produced mesons are $\pi^{0}$s or heavier mesons
which have main decay modes onto $\pi^{0}$s,
they produce high energy photons which are immediately 
converted to lower energy $\gamma$s
or $e^{\pm}$ pairs through $\gamma \rightarrow \gamma + \gamma$
or $\gamma \rightarrow e^{+} + e^{-}$ (Erber 1966)
for their large optical depth in strong magnetic fields.
However, if they are the charged mesons like $\pi^{\pm}$s, decay 
modes are very different and result in more interesting consequences.
Since the source emitters of charged nuclei are accelerated
to high energy (Hillas 1984) to some extent,
very high energy neutrinos are produced by 
$\pi^{+} \rightarrow \mu^{+} + \nu_{\mu} 
\rightarrow e^{+} + \nu_{e} + \bar{\nu_{\mu}} + \nu_{\mu}$
or its mirror conjugate process
for the  similar mechanism as proposed by Waxman \& Bahcall (1997).
These neutrinos could be a signature for strong meson emission 
from the GRB central engines and would be detected
in the very early phase of the short GRBs.

We formulate the meson synchrotron emission in the next section,
and the calculated results are shown and discussed in \S 3.

\section{Semi-Classical Treatment of Meson Emission}

We follow the established semi-classical treatment 
(Peskin \& Schroeder 1995, Itzykson \& Zuber 1980)
of synchrotron emission of the quantum fields interacting with 
classical source current in strong external fields.
In our present study the meson field is second quantized,
while the nucleons are not and obey classical motion.

The energy spectrum of $\pi^{0}$ synchrotron emission by 
a proton was first derived by 
Ginzburg \& Syrovatskii (1965a, b) and it is give by
\begin{equation}
\frac{dI_{\pi}}{dE_{\pi}} = \frac{g^{2}}{\sqrt{3} \pi}~
\frac{E_{\pi}}{\hbar^{2} c}~ \frac{1}{\gamma_{p}^{2}}~
\int_{y(x)}^{\infty} K_{1/3}(\eta)d\eta, \\
\end{equation}
where $g$ is the strong coupling constant, 
$g^{2}/\hbar c \approx 14$ (Sakurai 1967),  
$E_{\pi}$ is the energy of the emitted pion, and
$\gamma_{p}$ is the Lorentz factor,
$\gamma_{p} = E_{p}/ m_{p} c^{2}$, of the rotating proton
in a given magnetic field.  $K_{1/3}$ is the modified 
Bessel function of order $1/3$.
The function $y(x)$ is given by
\begin{equation}
y(x) = \frac{2}{3} \frac{m_{\pi}}{m_{p}} \frac{1}{\chi}
x \left(1 + \frac{1}{x^{2}} \right)^{3/2},
\end{equation}
where $m_{p} c^{2} = 938 MeV$ and $m_{\pi} c^{2} = 135 MeV$ are 
the rest masses of proton and $\pi^{0}$ meson, and 
the parameter $\chi$, which characterizes the synchrotron 
emission, is determined by the proton energy and the strength of 
magnetic field as $\chi = \frac{H}{H_{0}} \gamma_{p}$ with
$H_{0} \equiv \frac{m_{p}^{2} c^{3}}{e \hbar} 
= 1.5\times10^{20}G$.  
In the above equations, variable $x$ is introduced 
for mathematical simplicity, $x = \frac{E_{\pi}}{E_p} \times
\frac{m_p}{m_{\pi}}$.  Since the available energy of the pion 
satisfies $m_{\pi} \leq E_{\pi} \leq E_{p}$, 
the corresponding variable range 
of $x$ is $\gamma_{p}^{-1} \leq x \leq \frac{m_{p}}{m_{\pi}}$. 

Applying the same treatment to vector mesons, we extensively
obtain the energy spectrum of $\rho$ meson synchrotron emission 
\begin{equation}
\frac{dI_{\rho}}{dE_{\rho}} = \frac{g^{2}}{\sqrt{3} \pi}~
\frac{E_{\rho}}{\hbar^{2} c}~ \frac{1}{\gamma_{p}^{2}}~
\left( 1+ \frac{1}{x^{2}} \right)
\int_{y(x)}^{\infty} K_{5/3}(\eta)d\eta,
\end{equation}
where $K_{5/3}$ is the modified Bessel function of order $5/3$.
The function $y(x)$ is defined by Eq.~(2) by replacing $m_{\pi}$ with 
 $\rho$ meson mass $m_{\rho} c^{2} = 770 MeV$, and 
$x = \frac{E_{\rho}}{E_p} \times \frac{m_p}{m_{\rho}}$.  
Note that one can easily get the expression for 
photon synchrotron radiation in the limit of $m_{\rho} \rightarrow 0$
by replacing the strong coupling constant $g$ with 
the electromagnetic coupling constant $e$.

The total intensity of a scalar or vector meson 
as a function of $\chi$ is obtained by
integrating equation (1) or (3) over meson energies
$m_{\pi,\rho} \leq E_{\pi,\rho} \leq E_{p}$ or equivalently
over $\gamma_{p}^{-1} \leq x \leq 
\frac{m_{p}}{m_{\pi,\rho}}$.  It is useful to give an approximate
formula of the total intensity in the limit of large
or small $\chi$:
%
\begin{equation}
I_{\pi} = \left\{
\begin{array}{l@{\quad }l}
\frac{g^{2}}{6} \frac{m_{p}^{2}c^{3}}{\hbar^{2}},~~~
& {\rm :~~\chi~ \gg 1} \\
\frac{g^{2}}{\sqrt{3}} \frac{m_{\pi}m_{p}c^{3}}{\hbar^{2}}
    \chi \exp{ \left(-\frac{\sqrt{3}}{\chi}\frac{m_{\pi}}{m_{p}}\right)},~~~
& {\rm :~~\chi~ \ll 1}
\end{array}\right.
\end{equation}
%
and
\begin{equation}
I_{\rho} = \left\{
\begin{array}{l@{\quad }l}
\frac{27\sqrt[6]{3}}{16\pi}\Gamma(5/3)
\frac{2g^{2}}{3} \frac{m_{p}^{2}c^{3}}{\hbar^{2}} \chi^{2/3},~~~
&  :~~\chi~ \gg 1 \\
\frac{3}{2}\sqrt{\frac{3}{2}}
\left(\frac{m_p}{m_{\rho}}\right)
\left(1+ \left(\frac{m_{\rho}}{m_{p}} \right)^{2} \right)^{-1/4}
\left(2 \left(\frac{m_{\rho}}{m_{p}} \right)^{2} -1 \right)^{-1} \\
~~~~~~~~~ \times
\frac{g^{2}}{\sqrt{3}}\frac{m_{\rho} m_{p} c^{3}}{\hbar^{2}}
\chi^{3/2} \exp{\left(-\frac{2}{3 \chi}
\left(1+\left(\frac{m_{\rho}}{m_{p}}\right)^{2} \right)^{3/2}
\right)},~~~
& {\rm :~~\chi~ \ll 1}
\end{array}\right.
\end{equation}
where we have made an approximation
$K_{\nu}(\eta) \approx \frac{2^{\nu-1}\Gamma(\nu)}{\eta^{\nu}}$
for $\eta \ll 1$ ($\chi \gg 1$), or
$K_{\nu}(\eta) \approx \sqrt{\frac{\pi}{2\eta}}\exp{(-\eta)}$
for $\eta \gg 1$ ($\chi \ll 1$).
These approximations are in reasonable agreement with exact 
numerical integrals within $\pm$ 3\% for $\pi$ meson
and $\pm$ 10\% for $\rho$ meson at $\chi\leq0.01$
or $10^{2}\leq\chi$.

Let us make a short remark on our classical treatment.  
When the proton energy is very high or the external magnetic field 
is strong, i.e. $\chi \gg 1$, the quantum effects not only in the 
meson field but in the source nucleon current may not be negligible.
In the case of photon synchrotron radiation, 
quantum effects were carefully studied 
(Erber 1966), and semi-classical treatment was found to be 
a good approximation to the exact solution within a few percent.
In our treatment, the prefactor in Eq. (5) 
$\left(\frac{27\sqrt[6]{3}}{16\pi}\Gamma(5/3)\right) \approx 0.583$.
Taking the limit of $m_{\rho} \rightarrow 0$ and $g = e$,
Eq. (5) is applied to photon synchrotron radiation.
It is shown (Erber 1966) that this factor should be 0.5563 
in fully quantum mechanical calculation.  It therefore is expected 
to hold true for the hadron processes as well.

\section{Results and Discussions}

Figure 1a shows a comparison between the calculated spectra of 
scalar $\pi^{0}$ meson emission and $\gamma$ synchrotron radiation.
Since Usov (1992), Klu\'{z}niak \& Ruderman (1998), and 
Pacsy\'{n}ski (1998) suggested
strong magnetic fields of order $H\sim10^{16\sim17}G$, which are
presumed to associate with neutron stars or black holes of 
the GRB central engines, we here take $H=1.5\times10^{16}G$.  
The observed Lorentz factor of the fireball is $\Gamma\sim10^{3}$, 
which indicates that the energy of charged particles is at least
$\sim10^{12}eV$.  Although the acceleration mechanism 
in GRBs is still unknown, there are suggestions
(Milgrom \& Usov 1995) that the GRBs associate 
with UHECRs.  Six events of UHECRs beyond the Greisen-Zatsepin-Kuz'min
cutoff energy $\sim10^{20} eV$ (Hill \& Schramm 1985) have been 
detected by AGASA group (Takeda et al. 1998).
We therefore vary the proton energy from $E_{p}=10^{12}eV$ to $10^{22}eV$.
All calculated spectra cut off sharply at incident proton energy
$E_{\pi,\gamma} = E_{p}$.
As seen in this figure, the intensity of $\pi^{0}$ emission for 
$E_{p}=10^{12},10^{14}$ and $10^{16}eV$, which correspond 
respectively to $\chi\approx0.1,10$ and $1000$, 
is $10^{3}\sim10$ times stronger than that of $\gamma$ radiation 
in high energy parts of the spectra. 
This reflects the different coupling constants,
$g^{2}/e^{2}\sim10^{3}$. 
Very sharp declination of the $\pi^{0}$ spectra at lower energy  
arises from the integral in Eq. (1) for finite pion mass
because $K_{\nu}(\eta) \approx \sqrt{\frac{\pi}{2\eta}}\exp{(-\eta)}$
for $\eta \gg 1$, which is a good approximation in this energy region 
where $y(x) \gg 1$.

Figure 1b shows comparison between the calculated spectra of
$\rho$ meson emission and $\gamma$ synchrotron radiation.
In this figure both spectra look very similar to each other,
except for the absolute intensity and the sharp declination
of low energy spectra due to finite $\rho$ meson mass.
This is because the interactions between $\rho$ meson and proton 
and between photon and proton are of vector type.
For the proton energies above $\sim10^{14}eV$ the intensity of $\rho$ 
meson emission is roughly a thousand times larger than 
that of $\gamma$ synchrotron radiation, reflecting again
its stronger coupling constant.

Figure 2 displays the calculated total intensities of synchrotron 
emission of $\pi^{0}$ and $\rho$ mesons 
as a function of $\chi$.   These are the integrated spectra shown in 
Figs. 1a \& 1b over available meson energies.
The total intensity of $\gamma$ synchrotron radiation 
is also shown in this figure. 
$I_{\pi^{0}}$ exceeds $I_{\gamma}$
at $3\times10^{-2}<\chi<3\times10^{4}$, and  
$I_{\rho}$ exceeds $I_{\gamma}$ at $\chi>0.2$. 
Both $I_{\pi^{0}}$ and $I_{\rho}$ decrease exponentially 
with decreasing $\chi$ due to their finite masses.
(See the asymptotic forms at $\chi \ll 1$ in Eqs. (4) and (5).)
This sharp declination of $I_{\rho}$ takes place at higher $\chi$
than $I_{\pi^{0}}$ because $\rho$ meson mass is larger than pion mass.
$I_{\rho}$ resembles $I_{\gamma}$ at $\chi > 1$, except that
the intensity is a thousand times different from each other, 
reflecting that both $\rho$ meson and photon have the same 
vector type coupling to proton with different coupling 
constatnts, $g^{2}/e^{2}\sim10^{3}$.

A nucleon strongly couples to $\pi^{\pm}$ as well as $\pi^{0}$ field.
The interaction Hamiltonian is charge independent and
$H_{\rm int} = ig( \sqrt{2} \bar{\psi_{n}} \gamma_{5} \psi_{p} \phi_{\pi^{+}}
+ \bar{\psi_{p}} \gamma_{5} \psi_{p} \phi_{\pi^{0}})~ + c.c.$ 
(Sakurai 1967).  The initial state $\psi_{p}$ in both 
$p \rightarrow n + \pi^{+}$ and $p \rightarrow p + \pi^{0}$ 
processes is identical.  The difference comes from the final states:
Proton and $\pi^{+}$ couple to external magnetic field, but 
neutron and $\pi^{0}$ do not.  (Neutron has too small magnetic moment.) 
When we describe these processes in quantum mechanics, we should 
use wave functions from the solution of Dirac and Klein-Gordon equations
for the nucleon and the pion separately.
However, final states are more or less 
the same for the same charge state, aside from the different masses.
These might not change the reaction amplitiudes by many orders
at ultra-relativistic energies where the rest mass is neglected.
Note that the conjugate processes $n \rightarrow p + \pi^{-}$ and
$n \rightarrow n + \pi^{0}$ can also occur when a composite
nucleus like $^{56}Fe$ orbits in the strong magnetic fields.

Heavy mesons including $\rho$ meson have several appreciable 
branching ratios for the decay onto $\pi^{\pm}$.
Let us discuss what kinds of observational signals they may make.  

We assume that some fraction of $10^{51} \sim 10^{53} ergs$ 
of the gravitational or 
magnetic field energy is released by some unknown mechanism operating 
at the GRB central engine during very short time duration, $\sim 0.1ms$,
of the first sub-burst. We also assume that an appreciable part of this
energy is deposited into the relativistic motion of the material leading to
UHECRs.
In a somewhat different context, Waxman \& Bahcall (1997) proposed 
that the photomeson production in the ejecta of the fireball
would make a burst of $\sim10^{14}eV$ neutrinos.     
Although their proposed mechanism of meson production
is completely different from ours, we can apply similar discussion
on the physical consequence of  $\pi^{\pm}$ decay. 
Extending our previous discussions of $\pi^{0}$ in Fig. 2 to $\pi^{\pm}$,
we expect that a thousand times stronger intensity of
high energy neutrinos than the photons can be produced universally at 
$0.1 \leq \chi$ from $\pi^{+} \rightarrow \mu^{+} + \nu_{\mu} 
\rightarrow e^{+} + \nu_{e} + \bar{\nu_{\mu}} + \nu_{\mu}$
and its mirror conjugate process.
Since neutrinos can escape from the ambient matter, 
these neutrinos could be a clear signature showing strong
meson synchrotron emission near the central engines of GRBs 
associated with extremely strong magnetic fields.

The neutron emerging from $p \rightarrow n + \pi^{+}$
inherits almost all proton energy and can escape from
the region of strong magnetic field.  If there is not a dense shell
surrounding the central engine, it travels
$\sim10^{5}\times(\gamma_{n}/10^{10})~pc$ before beta decay.
This process may also produce a very high energy neutrino.
The generic picture of GRBs (M\'{e}sz\'{a}ros \& Rees 1993, Piran 1999) 
suggets that 
a baryon mass $\sim 10^{-5} M_{\odot}$ is involved in a single explosion. 
If this is the case, such amount of baryon mass is huge enough to stop 
almost all neutrons before running through the ambient matter.

If the central engines are neutron stars or black holes,
the material ejected from these compact objects contains heavy nuclei
such as oxygen and iron because these are the products from evolved 
massive stars.  Therefore, meson emission from a heavy nucleus as well as 
from a proton is worth being considered.
Let us consider a nucleus of total energy $E_{\rm tot}$, mass number A, 
and charge $Z$, in a magnetic field of strength $H$. 
The energy of each nucleon is $E = E_{\rm tot}/A$. 
When the strength of the effective magnetic field is 
$H_{\rm eff} = \frac{Z}{A} H$, the orbital trajectory of a proton
is the same as the trajectory of the nucleus in the magnetic field $H$.  
The intensity of $\pi^{0}$ emission by the nucleus should be the sum of 
each nucleonic contribution provided that the synchrotron 
emission is incoherent.  Thus the total intensity is given by
\begin{equation}
I_{\pi^{0}}^{(\rm A)}(E_{\rm tot},H) \approx  
A \times I_{\pi^{0}}^{(\rm p)}(E, H_{\rm eff}),
\end{equation}
where $I_{\pi^{0}}^{(\rm p)}(E, H_{\rm eff})$ is the intensity of $\pi^{0}$
emission by the nucleon of energy $E$ in magnetic field 
$H_{\rm eff}~$.  Note that both proton and neutron emit $\pi^{0}$. 
Figure 3 shows $I_{\pi^{0}}^{(\rm 56)}(E_{\rm tot},H)$ for $^{56}Fe$
as a function of total energy $E_{\rm tot}$
with a fixed magnetic field of $H=1.5\times10^{12}G$.
The sharp declination of $I_{\pi^{0}}^{(\rm 56)}$ takes place
at higher energy than $I_{\pi^{0}}^{(\rm p)}$ 
because each nucleon in $^{56}Fe$ has effectively smaller energy 
than a single proton.

There is now more motivation to study the GRBs in association with UHECRs.
It is highly desirable to proceed with a project like Orbiting Wide-angle 
Light-collectors (OWL),
in order to detect ultra-relativistic neutrinos from GRBs for
finding the true nature of the central engines.

This work has been supported in part by the Grant-in-Aid for Scientific 
Research (10640236, 10044103,11127220) of the Ministry of Education, Science,
Sports and Culture of Japan and also by JSPS-NSF Grant of the 
Japan-U.S. Joint Research Project.
We thank the Institute for Nuclear Theory at the University of Washington
for its hospitality and the Department of Energy for partial support during 
the completion of this work.

\section{Reference}
\noindent
Bhat, P.N., et al. 1992, Nature, 359, 217. \\ 
Erber, T. 1966, Rev. Mod. Phys. 38, 626. \\
Fishman, G.J., \& Meegan, C.A. 1995, 
  Ann. Rev. Astron. Astrophys. 33, 415. \\
Ginzburg, V.L., \& Syrovatskii, S.I. 1965a, Uspekhi Fiz. Nauk. 87, 65. \\
Ginzburg, V.L., \& Syrovatskii, S.I. 1965b, 
  Ann. Rev. Astron. Astrophys. 3, 297. \\
Hill, C.T., \& Schramm, D.N. 1985, Phys. Rev. D31, 564.  \\
Hillas, A.M. 1984, Ann. Rev. Astron. Astrophys., 22, 425. \\
Itzykson, C. \& Zuber, J.-B. 1980, Quantum Field Theory (McGraw-Hill, Inc.). \\ 
Kouveliotou, C. et al. 1998, Nature, 393, 235. \\
Klu\'{z}niak, W., \& Ruderman, M. 1998, Astrophs. J., 505, L113.  \\
M\'{e}sz\'{a}ros, P., \& Rees, M.J. 1993, Astrophys. J. 405, 278.  \\ 
Milgrom, M., \& Usov, V. 1995, Astrophs. J. , 449, L37. \\
Pacsy\'{n}ski, B. 1998, Astrophs. J., 494, L45. \\
Pacsy\'{n}ski, B., \& Xu, G. 1994, Astrophs. J., 427, 708. \\
Piran, T. 1999, Phys. Rep. 314, 575.  \\ 
Sari, R., \& Piran, T. 1997, Astrophys. J. 285, 270.  \\ 
Peskin, M.E., \& Schroeder, D.V. 1995, An Introduction to Quantum Field Theory
  (Addison-Wesley Publishing Company, Inc.). \\
Rees, M.J., \& M\'{e}sz\'{a}ros, P. 1992, Mon. Not. R. Astron. Soc. 258, 41P.  \\
Sakurai, J.J. 1967, Advanced Quantum Mechanics (Addison-Wesley Publishing
  Company, Inc.). \\
Takeda, M., et al. 1998, Phys. Rev. Lett. 81, 1163.  \\ 
Usov, V. 1992, Nature, 357, 472. \\
Waxman, E., \& Bahcall, J.N. 1997, Phys. Rev. Lett. 78, 2292. \\

\section{Figure Captions}

Figure 1:  (a) Calculated energy spectra of scalar $\pi^{0}$ meson 
emission (solid curve) and photon synchrotron radiation (dashed curve)
for various incident proton energies $E_{p}$ with a fixed magnetic 
field of $H=1.5\times10^{16}G$.  Denoted numbers in the figure 
are the proton energies $E_{p}=10^{12},10^{14},10^{16},10^{18},10^{20},$ and
$10^{22}eV$ from left to right.  
(b) The same as those in (a) for vector $\rho$ meson emission. 

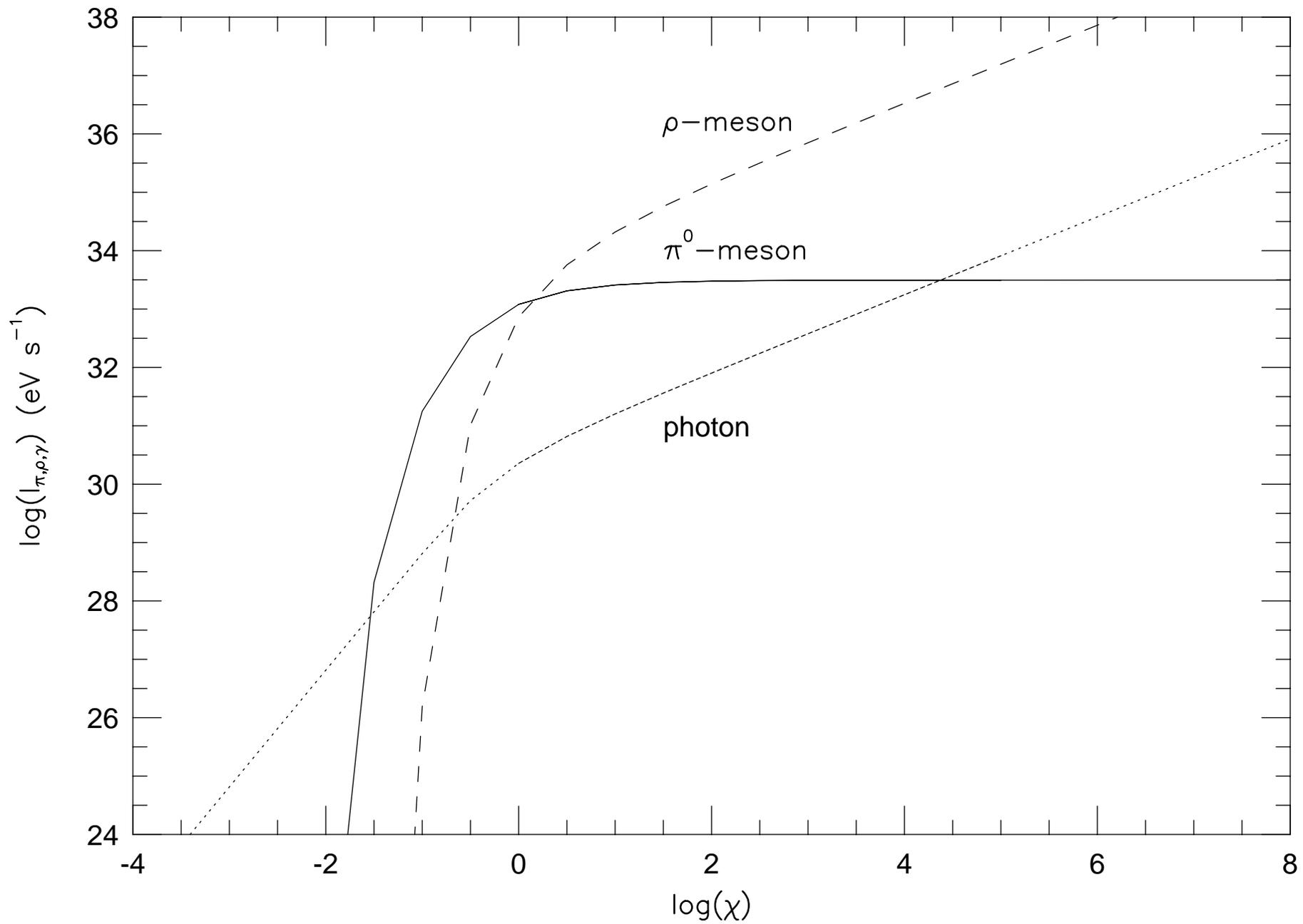
Figure 2: Calculated total intensities of the emission of scalar
$\pi^{0}$ meson (solid curve), vector $\rho$ meson 
(long-dashed curve), and photon $\gamma$ (dashed curve)
as a function of $\chi = \frac{H}{H_{0}} \gamma_{p}$.

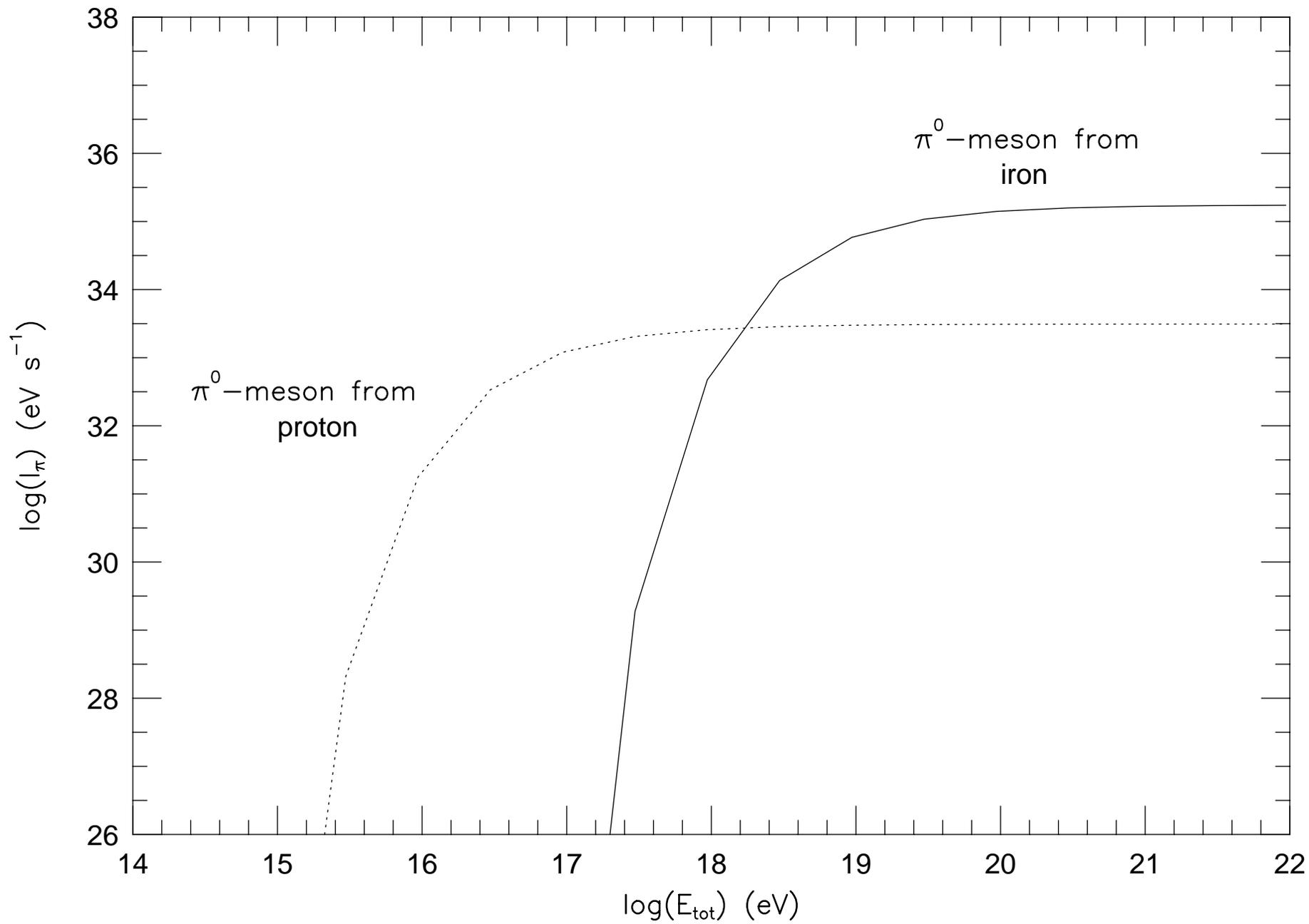
Figure 3: Calculated total intensities of scalar $\pi^{0}$ meson 
emission by the proton (dashed curve) and iron nucleus (solid curve) 
as a function of the total energy $E_{tot}$
with a fixed magnetic field of $H=1.5\times10^{12}G$.

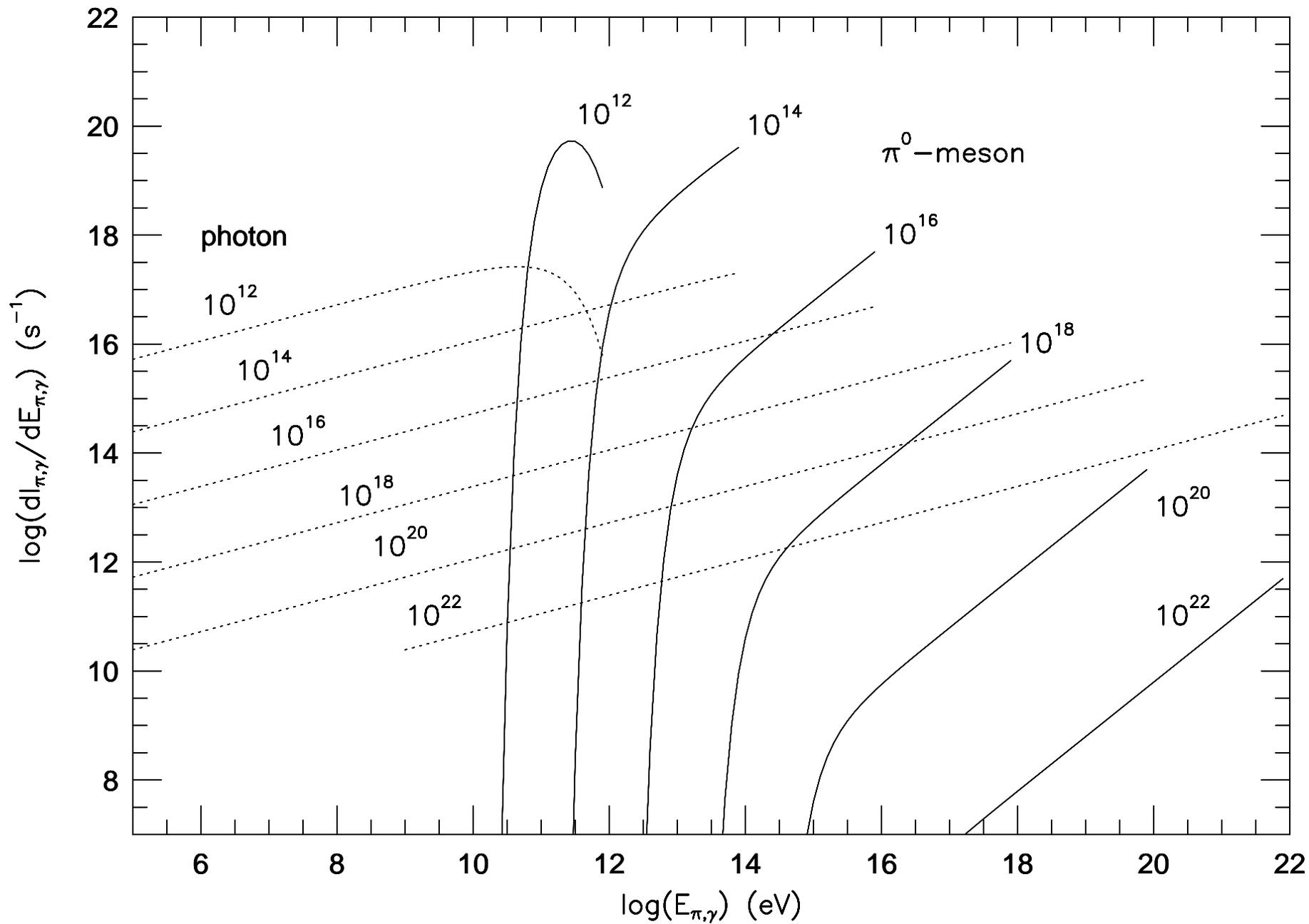
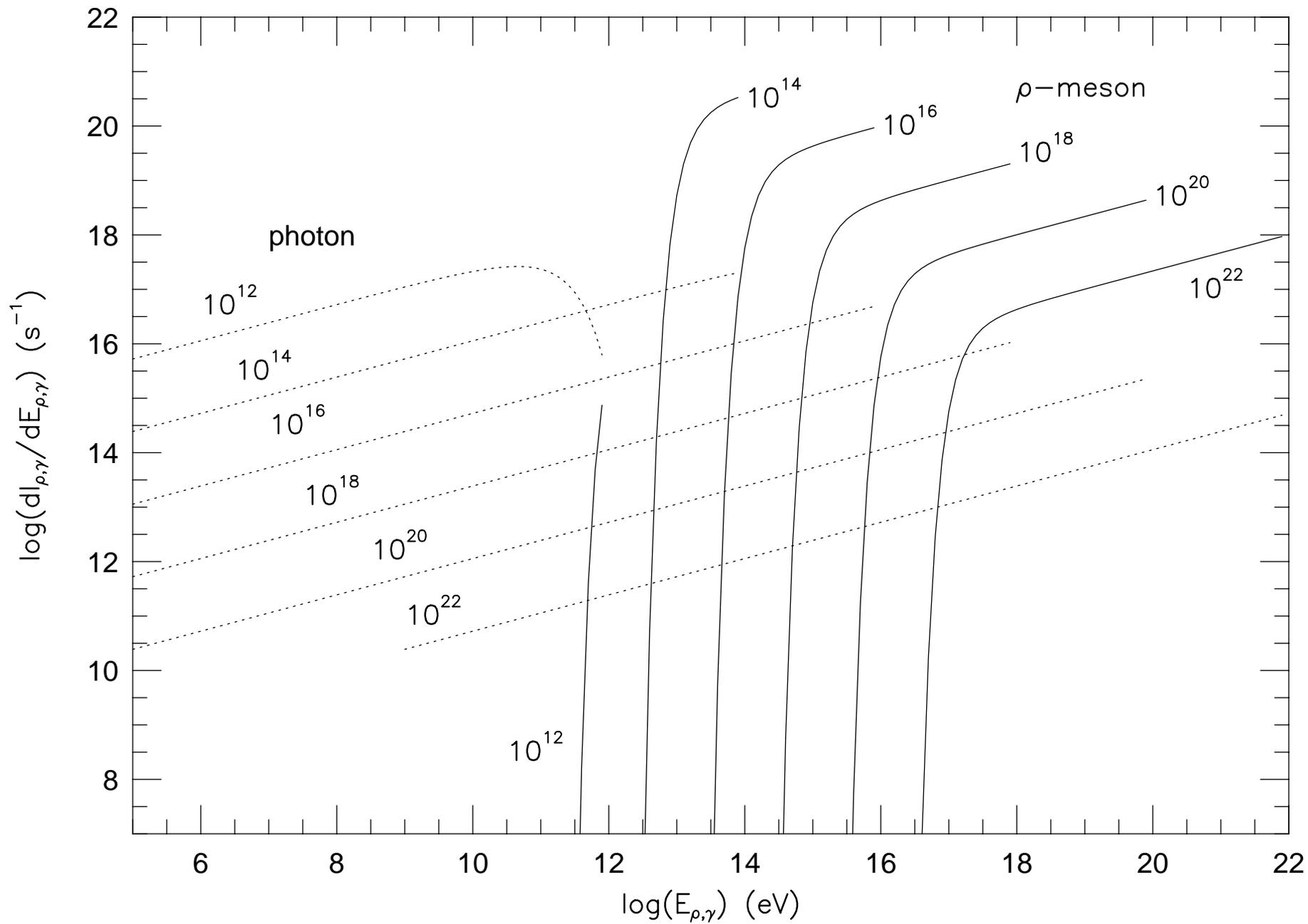
\end{document}